\documentclass{article}
\usepackage{slashed,verbatim,subfigure}
\usepackage[numbers,sort&compress]{natbib}
\usepackage{amsmath}
\usepackage{amssymb}
\usepackage{appendix}
\usepackage{graphicx}
\usepackage{dcolumn}
\usepackage{bm}
\usepackage[colorlinks=true,linktocpage=true,
linkcolor=blue,citecolor=blue]{hyperref}
\usepackage[a4paper]{geometry}
\newcommand{\be}{\begin{eqnarray}}
\newcommand{\ee}{\end{eqnarray}}
\newcommand{\ec}{\sigma_{\rm el}}
\newcommand{\eh}{\sigma_{\rm H}}
\newcommand{\ek}{\kappa_{\rm H}}
\begin{document}
\large
\title{\bf{Impact of nonextensivity on the transport coefficients of a magnetized hot and dense QCD matter}}
\author{Shubhalaxmi Rath\footnote{shubhalaxmi@iitb.ac.in}~~and~~Sadhana Dash\footnote{sadhana@phy.iitb.ac.in}\vspace{0.03in} \\ Department of Physics, Indian Institute of Technology Bombay, Mumbai 400076, India}
\date{}
\maketitle

\begin{abstract}
We have studied the impact of the nonextensivity on the transport coefficients related to charge and heat in thermal QCD. For this purpose, the electrical ($\sigma_{\rm el}$), Hall ($\sigma_{\rm H}$), thermal ($\kappa$) and Hall-type thermal ($\kappa_{\rm H}$) conductivities are determined using the kinetic theory approach in association with the nonextensive Tsallis statistical mechanism. The effect of nonextensivity is encoded in the nonextensive Tsallis distribution function, where the deviation of the parameter $q$ from 1 signifies the degree of nonextensivity in the concerned system. The thermal and electrical conductivities are found to increase with the introduction of nonextensivity, which means that the deviation of the medium from thermal equilibrium enhances both charge and heat transports. With the magnetic field, the deviations of $\sigma_{\rm el}$, $\sigma_{\rm H}$, $\kappa$ and $\kappa_{\rm H}$ from their respective equilibrated values increase, whereas these deviations decrease with the chemical potential. We have also studied how the extent of the nonextensivity modulates the longevity of magnetic field. Present work is further extended to the study of some observables associated with the aforesaid transport phenomena, such as the Knudsen number and the elliptic flow within the nonextensive Tsallis framework. 

\end{abstract}

\newpage

\section{Introduction}
The collisions of ultrarelativistic heavy ions at Relativistic Heavy Ion Collider (RHIC) and Large Hadron Collider (LHC) provide strong evidences for the production of a deconfined state of quarks and gluons, known as quark-gluon plasma (QGP). This extreme state of matter can be achieved at high temperature and/or high density. In addition, for noncentral events, strong magnetic fields are produced, whose strengths can vary between $eB=m_{\pi}^2$ ($\simeq 10^{18}$ Gauss) at RHIC and $eB=15$ $m_{\pi}^2$ at LHC \cite{Skokov:IJMPA24'2009,Bzdak:PLB710'2012}. These magnetic fields become weak with time, where the energy scale associated with the temperature prevails over the energy scale related to the magnetic field. However, the electrically conducting medium ensures the longevity of such magnetic fields by the Lenz's law \cite{Tuchin:AHEP2013'2013,McLerran:NPA929'2014,Rath:PRD100'2019}. Some of the phenomena induced due to the presence of the magnetic field are the chiral magnetic effect \cite{Fukushima:PRD78'2008,Kharzeev:NPA803'2008}, the axial magnetic effect \cite{Braguta:PRD89'2014,Chernodub:PRB89'2014}, the nonlinear electromagnetic current \cite{Kharzeev:PPNP75'2014,Satow:PRD90'2014}, the axial Hall current \cite{Pu:PRD91'2015}, the chiral vortical effect \cite{Kharzeev:PRL106'2011} etc. In recent years, the effects of magnetic field on QCD matter have been extensively studied, such as the thermodynamic and magnetic properties \cite{Rath:JHEP1712'2017,Bandyopadhyay:PRD100'2019,Rath:EPJA55'2019,
Karmakar:PRD99'2019}, the conductive properties \cite{Hattori:PRD94'2016,Feng:PRD96'2017,Fukushima:PRL120'2018,
Rath:PRD100'2019,Rath:EPJC80'2020,Thakur:PRD100'2019,Kurian:EPJC79'2019,Rath:EPJA59'2023}, the viscous properties \cite{Nam:PRD87'2013,Hattori:PRD96'2017,Li:PRD97'2018,Denicol:PRD98'2018,Das:PRD100'2019,
Rath:PRD102'2020,Rath:EPJC81'2021,Rath:EPJC82'2022}, the photon and dilepton productions from QGP \cite{Hees:PRC84'2011,Shen:PRC89'2014,Tuchin:PRC88'2013,Mamo:JHEP1308'2013}, the 
heavy quark diffusion \cite{Fukushima:PRD93'2016}, the magnetohydrodynamics \cite{Roy:PLB750'2015,Inghirami:EPJC76'2016} etc. In these studies, the nonextensive approach has not been followed. In high energy physics community, the Tsallis nonextensive distribution has recently achieved great importance as it shows good fits of the transverse momentum distributions for a wide range of collision energies by STAR \cite{Abelev:PRC75'2007}, PHENIX \cite{Adare:PRC83'2011}, ALICE \cite{Aamodt:EPJC71'2011} and CMS \cite{Khachatryan:JHEP1105'2011} collaborations. The nonextensive Tsallis statistics is considered as a generalization of the Boltzmann-Gibbs statistics, where the parameter $q$ that measures the extent of non-equilibration is called the nonextensive Tsallis parameter. The theory groups have also started considering the nonextensive statistics in the study of thermodynamics, transport processes etc. \cite{Tsallis:JSP52'1988,Tsallis:BOOK'2009,Tsallis:EPJA40'2009,Beck:EPJA40'2009,
Kaniadakis:EPJA40'2009,Kodama:EPJA40'2009,Wilk:EPJA40'2009,Alberico:EPJA40'2009,
Biro:EPJA40'2009,Alqahtani:EPJC82'2022}. In Langevin, Fokker-Planck, and/or Boltzmann type equations, this deviation can be incorporated through a nonextensive parameter or Tsallis parameter, $q$, where $q=1$ represents the Boltzmann limit \cite{Walton:PRL84'2000,Kaniadakis:PLA288'2001,Biro:PRL94'2005,Sherman:LNP633'2004}. The relation between the parameter $q$ and temperature fluctuations has been studied in ref. \cite{Cleymans:JPG36'2009}, which observed that the deviation of $q$ from unity measures the fluctuation in the temperature and there is no temperature fluctuation in the Boltzmann limit ($q=1$). 

The fits to the RHIC and LHC spectra suggest that $q$ for the hadronic matter can deviate up to 1.08 - 1.2 \cite{Tang:PRC79'2009,Shao:JPG37'2010,Sikler:EPJWC13'2011} and for a quark matter, the value of $q$ deviates up to 1.22 \cite{Biro:JPG36'2009}. The value of $q$ should thus be considered as never being far from 1. Similar result for the parameter $q$ has also been obtained in an analysis of the composition of final-state particles \cite{Cleymans:JPG36'2009}. For large $p_T$ results of particle production with $q>1$, the freeze-out temperature becomes smaller in order to keep the particle yields the same, whereas, to compensate the decrease in the particle number, the baryon chemical potential increases with $q>1$ \cite{Cleymans:JPG36'2009}. Some observation shows that the Tsallis distribution leads to a much better chemical equilibrium than the Boltzmann distribution with $q=1$ \cite{Cleymans:JPG36'2009}. The nonextensivity can be inserted in the underlying dynamical model through the Tsallis nonextensive statistics. It is very beneficial to use the nonextensive Tsallis distribution function in the dynamical model itself to describe the characteristics of the medium in both the QGP and the freeze-out phases in order to study the effects on the bulk observables. The observables in heavy ion collisions, such as the transverse momentum spectra, the multiplicity fluctuations, the nuclear modification factor etc. are influenced by the nonextensive parameter $q$ \cite{Biro:JPG35'2008,Beck:PA322'2003,Tripathy:EPJA52'2016}. Thus, when the value of the nonextensive parameter gets deviated from unity, it is highly expected to leave some noticeable impacts on the properties of the matter produced in heavy ion collisions. Our present work attempts to see the possible deviation of the charge and heat transport properties of the QCD medium when $q$ is slightly above unity. 

In this paper, the transport coefficients related to charge conduction and heat conduction are calculated for the first time using a nonextensive relativistic Boltzmann transport equation in the relaxation time approximation, where the nonextensive Tsallis formalism has been incorporated. In addition, the thermal masses of particles and the weak magnetic field limit ($T^2 \gg |q_fB|$, where $|q_f|$ is the absolute electronic charge of the quark with flavor $f$) are considered in the calculations. This study is relevant to understand the effect of the nonextensivity on the transport coefficients in hot QCD matter. The effect on the lifetime of magnetic field is also studied by varying the value of parameter $q$ up to 1.2. The aforesaid transport coefficients are important to understand the local equilibrium property, elliptic flow, hydrodynamic evolution of the strongly interacting matter etc. Thus, the deviation if any, of these transport coefficients due to the nonextensivity could also leave significant imprints on the observables used at heavy ion collisions. 

The present paper is organized as follows. Section 2 is dedicated to the study 
of different charge and heat conductivities of a weakly magnetized QCD medium 
assuming a nonextensive scenario within the kinetic theory approach and its 
effect on the lifetime of magnetic field. The results on different conductivities 
are discussed in section 3. In section 4, some observables related to the 
aforesaid transport phenomena are studied. Section 5 presents the conclusions of this work. 

\section{Charge and heat conductivities in the nonextensive Tsallis framework}
This section is devoted to the calculation of different charge and heat conductivities by 
using the relativistic Boltzmann transport equation in the relaxation time approximation, 
where the effect of nonextensivity is incorporated through the Tsallis distribution 
function. In particular, subsection 2.1 contains the calculation of the conductivities at 
zero magnetic field, subsection 2.2 shows the effect of nonextensivity on the lifetime of 
magnetic field and in subsection 2.3, the conductivities are determined in a weak magnetic 
field. 

\subsection{Hot and dense QCD matter in the absence of magnetic field}
In this subsection, we study the response of the nonextensivity to the charge and heat flow by calculating the electrical and thermal conductivities in the nonextensive Tsallis framework in the absence of magnetic field. 

\subsubsection{Response of the nonextensivity to the charge flow in a QCD medium}
In the nonextensive Tsallis framework, the fermion distribution function is represented \cite{Conroy:PRD78'2008,Biro:PRC85'2012,Cleymans:JPG39'2012} as
\begin{eqnarray}\label{N.D.F.}
f_q=\frac{1}{\left[1+(q-1)\beta\omega\right]^{\frac{1}{q-1}}+1}, ~ {\rm with} ~ \beta\omega> 0
,\end{eqnarray}
where $\omega=\omega_f-\mu_f$ for quarks and $\bar{\omega}=\omega_f+\mu_f$ for antiquarks, and $q$ 
represents the nonextensive parameter. Here, $\omega_f=\sqrt{\mathbf{p}^2+m_f^2}$ and $\mu_f$ is the chemical potential of quark with flavor $f$. The deviation of $q$ from unity signifies the nonextensivity of the system. At high temperatures, Fermi-Dirac statistics (for fermions) and Bose-Einstein statistics (for bosons) behave like Boltzmann statistics, so, the above distribution function can be approximated to 
\begin{eqnarray}\label{N.D.F.(A.)}
f_q\approx\left[1+(q-1)\beta\omega\right]^{\frac{1}{1-q}}
~,\end{eqnarray}
which is also the exponential factor in the usual Fermi-Dirac distribution function. Expansion of eq. \eqref{N.D.F.(A.)} around $q=1$ gives
\begin{eqnarray}\label{N.D.F.(A.E.)}
f_q=e^{-\beta\omega}+\frac{1}{2}(q-1)\beta^2\omega^2e^{-\beta\omega}+\frac{1}{24}(q-1)^2(3\beta\omega-8)\beta^3\omega^3e^{-\beta\omega}+...
~,\end{eqnarray}
where the first term in the right hand side represents the Boltzmann distribution function. One can see that at $q=1$, only leading term remains, {\em i.e.} $f_q=e^{-\beta\omega}$. Thus, the limit $q\rightarrow1$ yields the Boltzmann distribution and the deviation of $q$ from unity explains how much the medium drives away from the equilibrated thermal distribution of particles. Let us consider a nonextensive QCD medium having three flavors ($f=u,d,s$). When this medium comes under the effect of an external electric field, an electric current density is induced whose spatial component can be written as
\begin{eqnarray}\label{current}
J^i = \sum_f g_f \int\frac{d^3\rm{p}}{(2\pi)^3\omega_f}
p^i [q_f\delta f_q+{\bar q_f}\delta \bar{f_q}]
~,\end{eqnarray}
where $g_f$, $q_f$ ($\bar q_f$) and $\delta f_q$ ($\delta \bar{f_q}$) are the degeneracy factor, electric charge and infinitesimal change in the nonextensive Tsallis distribution function for the quark (antiquark) of $f$th flavor, respectively. The Ohm's law states that the spatial current density is proportional to the electric field, with the proportionality factor being the electrical conductivity, {\em i.e.}, 
\begin{eqnarray}\label{Ohm's law (1)}
J^i=\ec E^i
~.\end{eqnarray}
In order to determine the infinitesimal shift $\delta f_q$, we use the relativistic Boltzmann transport equation (RBTE) in the relaxation time approximation (RTA) within the nonextensive Tsallis mechanism, {\em i.e.}, 
\be\label{R.B.T.E.1}
p^\mu\frac{\partial f_q^\prime}{\partial x^\mu}+q_f F^{\rho\sigma} 
p_\sigma \frac{\partial f_q^\prime}{\partial p^\rho}=-\frac{p_\nu u^\nu}{\tau_f}\delta f_q
~,\ee
where $f_q^\prime=\delta f_q+f_q$, $F^{\rho\sigma}$ denotes the electromagnetic field strength tensor whose components are associated with the electric and magnetic fields. In the above equation, the relaxation time for quarks (antiquarks), $\tau_f$ ($\tau_{\bar{f}}$) is given \cite{Hosoya:NPB250'1985} by
\begin{eqnarray}
\tau_{f(\bar{f})}=\frac{1}{5.1T\alpha_s^2\log\left(1/\alpha_s\right)\left[1+0.12(2N_f+1)\right]}
~.\end{eqnarray}
In the absence of magnetic field, in order to see the response of electric field, we use the components of $F^{\rho\sigma}$ related to only electric field. In addition, for a spatially homogeneous distribution function with the steady-state condition, one can use $\frac{\partial f_q^\prime}{\partial \mathbf{r}}=0$ and 
$\frac{\partial f_q^\prime}{\partial t}=0$. Thus, RBTE \eqref{R.B.T.E.1} takes the following form, 
\be\label{R.B.T.E.2}
q_f\mathbf{E}\cdot\mathbf{p}\frac{\partial f_q^\prime}{\partial p_0}
+q_f p_0\mathbf{E}\cdot\frac{\partial f_q^\prime}{\partial \mathbf{p}}
=-\frac{p_0}{\tau_f}\delta f_q
~.\ee
Solving eq. \eqref{R.B.T.E.2} with the nonextensive distribution function, 
we get $\delta f_q$ as
\be
\delta f_q=\frac{2\tau_fq_f\beta\mathbf{E}\cdot\mathbf{p}}{\omega_f}\left[1+(q-1)\beta(\omega_f-\mu_f)\right]^{\frac{q}{1-q}}
~.\ee
Similarly, $\delta \bar{f_q}$ is calculated as
\be
\delta \bar{f_q}=\frac{2\tau_{\bar{f}}{\bar q_f}\beta\mathbf{E}\cdot\mathbf{p}}{\omega_f}\left[1+(q-1)\beta(\omega_f+\mu_f)\right]^{\frac{q}{1-q}}
~.\ee
Using the values of $\delta f_q$ and $\delta \bar{f_q}$ in eq. (\ref{current}) 
and then comparing with eq. \eqref{Ohm's law (1)}, we get the electrical 
conductivity as
\be\label{I.E.C.}
\nonumber\sigma_{\rm el} &=& \frac{\beta}{3\pi^2}\sum_f g_f q_f^2\int d{\rm p}~\frac{{\rm p}^4}{\omega_f^2} ~ \left[\tau_f \lbrace1+(q-1)\beta(\omega_f-\mu_f)\rbrace^{\frac{q}{1-q}}\right. \\ && \left.+\tau_{\bar{f}}\lbrace1+(q-1)
\beta(\omega_f+\mu_f)\rbrace^{\frac{q}{1-q}}\right]
.\ee

\subsubsection{Response of the nonextensivity to the heat flow in a QCD medium}
The flow of heat in a medium is regulated by the temperature and pressure gradients and the corresponding heat flow four-vector is defined as
\be\label{heat flow (1)}
Q_\mu=\Delta_{\mu\alpha}T^{\alpha\beta}u_\beta-h\Delta_{\mu\alpha}N^\alpha
,\ee
where the first and second terms in the right hand side are distinguished as the energy diffusion and the enthalpy diffusion, respectively. Here, $T^{\alpha\beta}$ 
is the energy-momentum tensor, $N^\alpha$ is the particle flow four-vector, the 
projection operator $\Delta_{\mu\alpha}=g_{\mu\alpha}-u_\mu u_\alpha$, the enthalpy per particle $h=(\varepsilon+P)/n$ with $\varepsilon$, $P$ and $n$ denoting the energy 
density, the pressure and the particle number density, respectively. $N^\alpha$ and $T^{\alpha\beta}$ are the first and second moments of the nonextensive distribution function, 
respectively. 
\be
&&N^\alpha=\sum_f g_f\int \frac{d^3{\rm p}}{(2\pi)^3\omega_f}p^\alpha \left[f_q+\bar{f}_q\right] ~ \label{P.F.F.}, \\ 
&&T^{\alpha\beta}=\sum_f g_f\int \frac{d^3{\rm p}}{(2\pi)^3\omega_f}p^\alpha p^\beta \left[f_q+\bar{f}_q\right] ~ \label{E.M.T.}
.\ee
From $N^\alpha$ and $T^{\alpha\beta}$, one can get $n=N^\alpha u_\alpha$, $\varepsilon=u_\alpha T^{\alpha\beta} u_\beta$ and $P=-\Delta_{\alpha\beta}T^{\alpha\beta}/3$. In the rest 
frame of the heat bath, the heat flow is purely spatial, because it is orthogonal to the fluid 
four-velocity. Thus, the spatial component of heat flow is written as
\be\label{heat1 (1)}
Q^i=\sum_f g_f\int \frac{d^3{\rm p}}{(2\pi)^3} ~ \frac{p^i}{\omega_f}\left[(\omega_f-h_f)\delta f_q+(\omega_f-\bar{h}_f)\delta \bar{f}_q\right]
~.\ee
Through the Navier-Stokes equation, the heat flow is associated with the gradients 
of temperature and pressure as
\be\label{heat.1}
Q^i=-\kappa\delta^{ij}\left[\partial_j T - \frac{T}{\varepsilon+P}\partial_j P\right] 
,\ee
where $\kappa$ represents the thermal conductivity. For the calculation of thermal conductivity, the electromagnetic field strength part can be dropped from the relativistic Boltzmann transport 
equation (\ref{R.B.T.E.1}) and then expanding the gradient of the nonextensive distribution 
function in terms of the gradients of flow velocity and temperature, we have
\be\label{eq1.1}
-p^\mu\left[1+(q-1)\beta\omega\right]^{\frac{q}{1-q}}\left[\left(u_\alpha p^\alpha\right)\partial_\mu\beta+\beta\partial_\mu\left(u_\alpha p^\alpha\right)-\partial_\mu\left(\beta\mu\right)\right]=-\frac{u_\nu p^\nu}{\tau_f}\delta f_q
~.\ee
After solving eq. (\ref{eq1.1}), we get $\delta f_q$ as
\be\label{delta.q1}
\nonumber\delta f_q &=& -\beta\tau_f\left[1+(q-1)\beta\omega\right]^{\frac{q}{1-q}}\left[\frac{\left(\omega_f-h_f\right)}{T}v^j\left(\partial_jT-\frac{T}{nh_f}\partial_jP\right)\right. \\ && \left.+p_0\frac{DT}{T}-\frac{p^\mu p^\alpha}{p_0}\nabla_\mu u_\alpha+TD\left(\frac{\mu_f}{T}\right)\right]
.\ee
Similarly, $\delta \bar{f}_q$ is determined as
\be\label{delta.aq1}
\nonumber\delta \bar{f_q} &=& -\beta\tau_{\bar{f}}\left[1+(q-1)\beta\bar{\omega}\right]^{\frac{q}{1-q}}\left[\frac{\left(\omega_f-\bar{h}_f\right)}{T}v^j\left(\partial_jT-\frac{T}{n\bar{h}_f}\partial_jP\right)\right. \\ && \left.+p_0\frac{DT}{T}-\frac{p^\mu p^\alpha}{p_0}\nabla_\mu u_\alpha-TD\left(\frac{\mu_f}{T}\right)\right]
.\ee
Substituting $\delta f_q$ and $\delta \bar{f_q}$ in eq. (\ref{heat1 (1)}) and comparing 
with eq. (\ref{heat.1}), we get the thermal conductivity as
\be\label{I.T.C.}
\nonumber\kappa &=& \frac{\beta^2}{6\pi^2}\sum_fg_f\int d{\rm p} \frac{{\rm p}^4}{\omega_f^2} ~ \left[\tau_f(\omega_f-h_f)^2\lbrace1+(q-1)\beta(\omega_f-\mu_f)\rbrace^{\frac{q}{1-q}}\right. \\ && \left.+\tau_{\bar{f}}(\omega_f-\bar{h}_f)^2\lbrace1+(q-1)\beta(\omega_f+\mu_f)\rbrace^{\frac{q}{1-q}}\right]
.\ee

\subsection{Effect of the nonextensivity on the lifetime of magnetic field}
This subsection is dedicated to see how the nonextensivity affects the lifetime of magnetic field produced in the initial stages of the noncentral ultrarelativistic heavy ion collisions. Although this magnetic field is transient, but finite electrical conductivity elongates its lifetime significantly. This was discerned previously for a thermal distribution, where the nonextensive Tsallis parameter, $q=1$. Now, it is interesting to observe how the longevity of magnetic field gets influenced when the nonextensive Tsallis parameter becomes deviated from unity. 

\begin{figure}[]
\begin{center}
\includegraphics[width=13.99cm]{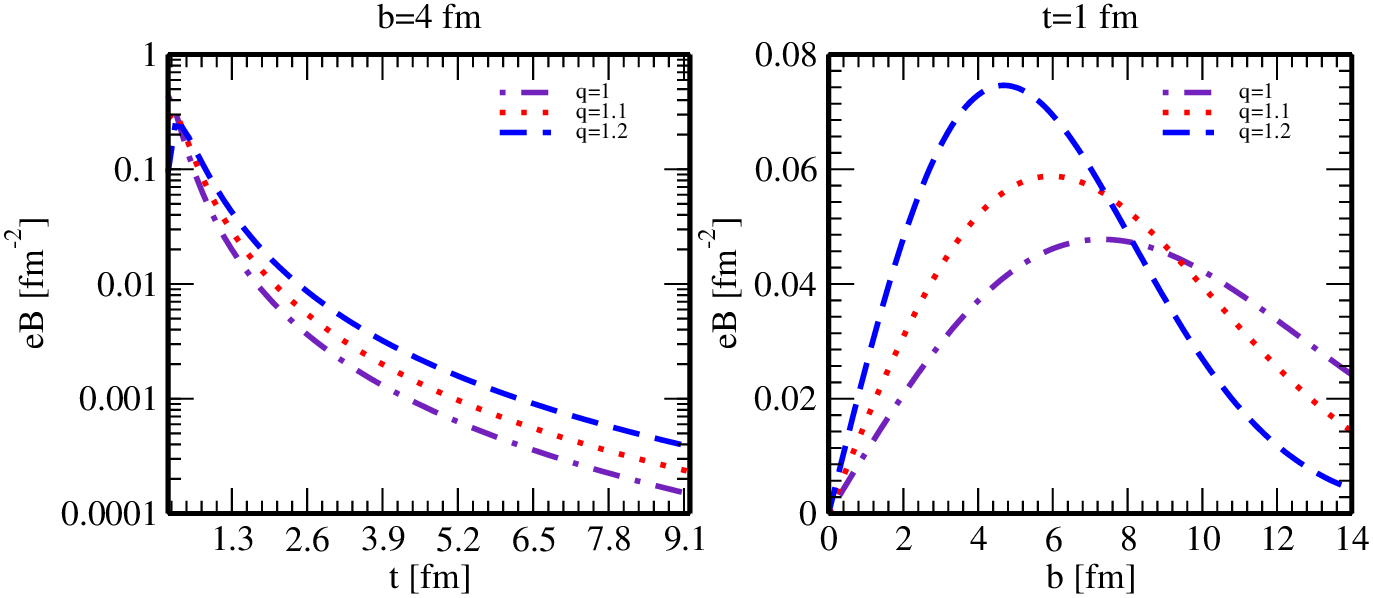}
\caption{Variation of magnetic field with time (left panel) and with impact parameter (right panel) at different values of the nonextensive parameter.}\label{eb}
\end{center}
\end{figure}

Let a charged particle moves along $x$-direction, then there will be a production of 
magnetic field transverse to the trajectory of particle, which can be expressed \cite{Tuchin:AHEP2013'2013} as
\be\label{eb1}
e\mathbf{B}_{\rm medium}=\frac{e^2b\sigma_{\rm el}}{8\pi(t-x)^2}e^{-\frac{b^2\sigma_{\rm el}}{4(t-x)}}\hat{\mathbf{z}}
~,\ee
where $b$ is the impact parameter. In eq. \eqref{eb1}, the electrical conductivity depends on the time through the cooling law, $T^3\propto{t^{-1}}$. By taking the initial time $0.2$ fm and initial temperature at $390$ MeV, figure \ref{eb} is plotted which shows the variations of magnetic field with time (left panel) for $x=0$, $b=4$ fm and with impact parameter (right panel) for $x=0$, $t=1$ fm in an electrically conducting medium at different $q$ values. 

It is well known that the magnetic field decays very fast in vacuum, whereas the 
electrically conducting thermal medium helps in delaying the decay of magnetic 
field, where the parameter $q$ has always been set to unity. The left panel of 
figure \ref{eb} displays that the decay of magnetic field gets further slowed 
down as the value of $q$ increases slightly above unity. Thus, the nonextensivity 
of the thermal medium extends the lifetime of a nearly stable magnetic field. So, 
it is pertinent to study how the nonextensivity modifies the properties of the hot 
QCD matter in the absence as well as in the presence of magnetic field. It is also 
observed that, at initial time, the effect of the nonextensivity on the magnetic 
field is meagre and the effect becomes more conspicuous with the increase of the time. 
The right panel of figure \ref{eb} shows the effect of the nonextensivity on the 
magnetic field with the increase of impact parameter. It shows peak at some specific combination of magnetic field and impact parameter, and the height of the peak 
increases with the increase of deviation of $q$ from unity, which occurs at a lower 
impact parameter. Thus, with the increase of the nonextensivity, the peak value of 
magnetic field increases. 

\subsection{Hot and dense QCD matter in a weak magnetic field}
In this subsection, we study the response of the nonextensivity to the charge and heat flow by calculating the electrical, Hall, thermal and Hall-type thermal conductivities in the nonextensive Tsallis framework 
in the presence of a weak magnetic field. 

\subsubsection{Response of the nonextensivity to the charge flow in a weakly magnetized QCD medium}
At finite magnetic field, the spatial component of electric current density is written as
\begin{eqnarray}\label{Multicomponent structure (1)}
J^i=\sigma_0\delta^{ij}E_j+\sigma_1\epsilon^{ijk}b_kE_j+\sigma_2b^ib^jE_j
~,\end{eqnarray}
where $\sigma_0$, $\sigma_1$ and $\sigma_2$ are various components of charge transport and $\mathbf{b}=\frac{\mathbf{B}}{B}$ is the direction of magnetic field. For $\mathbf{E}\perp\mathbf{B}$, the third term in the right hand side vanishes and thus, eq. \eqref{Multicomponent structure (1)} becomes
\begin{eqnarray}\label{Multicomponent structure}
J^i=\sigma^{ij}E_j=\left(\sigma_{\rm el}\delta^{ij}+\sigma_{\rm H}\epsilon^{ij}\right)E_j
~,\end{eqnarray}
where $\epsilon^{ij}$ is the antisymmetric $2\times2$ unit matrix, $\sigma_0=\sigma_{\rm el}$ denotes the electrical conductivity and $\sigma_1=\sigma_{\rm H}$ represents the Hall conductivity. To determine the infinitesimal change in the nonextensive distribution function at finite magnetic field, let us rewrite RBTE \eqref{R.B.T.E.1} as
\be\label{R.B.T.E.}
p^\mu\frac{\partial f_q^\prime}{\partial x^\mu}+\mathcal{F}^\mu\frac{\partial f_q^\prime}{\partial p^\mu}=-\frac{p_\nu u^\nu}{\tau_f}\delta f_q
~,\ee
where $\mathcal{F}^\mu=q_fF^{\mu\nu}p_\nu=(p^0\mathbf{v}\cdot\mathbf{F}, p^0\mathbf{F})$ and the Lorentz force is defined as $\mathbf{F}=q_f(\mathbf{E}+\mathbf{v}\times\mathbf{B})$. For a spatially homogeneous distribution function with the steady-state condition, eq. \eqref{R.B.T.E.} gets simplified into
\be\label{R.B.T.E.(1)}
\mathbf{v}\cdot\mathbf{F}\frac{\partial f_q^\prime}{\partial p_0}+\mathbf{F}\cdot\frac{\partial f_q^\prime}{\partial \mathbf{p}}=-\frac{(f_q^\prime-f_q)}{\tau_f}
~.\ee
For $\mathbf{E}=E\hat{x}$ and $\mathbf{B}=B\hat{z}$, we have
\be\label{R.B.T.E.(2)}
\tau_fq_fEv_x\frac{\partial f_q^\prime}{\partial p_0}+\tau_fq_fBv_y\frac{\partial f_q^\prime}{\partial p_x}-\tau_fq_fBv_x\frac{\partial f_q^\prime}{\partial p_y}=f_q-f_q^\prime-\tau_fq_fE\frac{\partial f_q}{\partial p_x}
~.\ee
To solve the above equation, we have followed an ansatz which is given \cite{Feng:PRD96'2017} by
\be\label{ansatz}
f_q^\prime=f_q-\tau_fq_f\mathbf{E}\cdot\frac{\partial f_q}{\partial \mathbf{p}}-\mathbf{\Gamma}\cdot\frac{\partial f_q}{\partial \mathbf{p}}
~,\ee
where the quantity $\mathbf{\Gamma}$ needs to be evaluated. Using ansatz \eqref{ansatz} in eq. \eqref{R.B.T.E.(2)} within the nonextensive framework, we get
\be\label{R.B.T.E.(3)}
\tau_fq_fEv_x\frac{\partial f_q^\prime}{\partial p_0}+\beta\left[1+(q-1)\beta\omega\right]^{\frac{q}{1-q}}\left(\Gamma_xv_x+\Gamma_yv_y+\Gamma_zv_z\right)-q_fB\tau_f\left(v_x\frac{\partial f_q^\prime}{\partial p_y}-v_y\frac{\partial f_q^\prime}{\partial p_x}\right)=0
~.\ee
The partial derivatives in the above equation are determined as follows, 
\be
\nonumber\frac{\partial f_q^\prime}{\partial p_0} &=& -\beta\left[1+(q-1)\beta\omega\right]^{\frac{q}{1-q}}-q_f\tau_fE\frac{\beta v_x}{\omega_f}\left[1+(q-1)\beta\omega\right]^{\frac{q}{1-q}} \\ && \nonumber -q_f\tau_fEv_x\beta^2q\left[1+(q-1)\beta\omega\right]^{\frac{2q-1}{1-q}} \\ && \nonumber -\frac{\beta}{\omega_f}\left[1+(q-1)\beta\omega\right]^{\frac{q}{1-q}}\left(\Gamma_xv_x+\Gamma_yv_y+\Gamma_zv_z\right) \\ && -\beta^2
q\left[1+(q-1)\beta\omega\right]^{\frac{2q-1}{1-q}}
\left(\Gamma_xv_x+\Gamma_yv_y+\Gamma_zv_z\right), \\ 
\nonumber\frac{\partial f_q^\prime}{\partial p_x} &=& -\beta v_x\left[1+(q-1)\beta\omega\right]^{\frac{q}{1-q}}+q_f\tau_fE\frac{\beta}{\omega_f}\left[1+(q-1)\beta\omega\right]^{\frac{q}{1-q}} \\ && \nonumber -q_f\tau_fE\frac{\beta v_x^2}{\omega_f}\left[1+(q-1)\beta\omega\right]^{\frac{q}{1-q}}-q_f\tau_fE\beta^2 v_x^2q\left[1+(q-1)\beta\omega\right]^{\frac{2q-1}{1-q}} \\ && \nonumber+\frac{\beta\Gamma_x}{\omega_f}\left[1+(q-1)\beta\omega\right]^{\frac{q}{1-q}}-\frac{\beta v_x^2\Gamma_x}{\omega_f}\left[1+(q-1)\beta\omega\right]^{\frac{q}{1-q}} \\ && \nonumber -\beta^2v_x^2q\Gamma_x\left[1+(q-1)\beta\omega\right]^{\frac{2q-1}{1-q}}-\frac{\beta v_x^2\Gamma_y}{\omega_f}\left[1+(q-1)\beta\omega\right]^{\frac{q}{1-q}} \\ && \nonumber -\beta^2v_xv_yq\Gamma_y\left[1+(q-1)\beta\omega\right]^{\frac{2q-1}{1-q}}-\frac{\beta v_xv_z\Gamma_z}{\omega_f}\left[1+(q-1)\beta\omega\right]^{\frac{q}{1-q}} \\ && -\beta^2v_xv_zq\Gamma_z\left[1+(q-1)\beta\omega\right]^{\frac{2q-1}{1-q}}, \\ 
\nonumber\frac{\partial f_q^\prime}{\partial p_y} &=& -\beta v_y\left[1+(q-1)\beta\omega\right]^{\frac{q}{1-q}}-q_f\tau_fE\frac{\beta v_xv_y}{\omega_f}\left[1+(q-1)\beta\omega\right]^{\frac{q}{1-q}} \\ && \nonumber -q_f\tau_fE\beta^2 v_xv_yq\left[1+(q-1)\beta\omega\right]^{\frac{2q-1}{1-q}}-\frac{\beta v_xv_y\Gamma_x}{\omega_f}\left[1+(q-1)\beta\omega\right]^{\frac{q}{1-q}} \\ && \nonumber -\beta^2v_xv_yq\Gamma_x\left[1+(q-1)\beta\omega\right]^{\frac{2q-1}{1-q}}+\frac{\beta\Gamma_y}{\omega_f}\left[1+(q-1)\beta\omega\right]^{\frac{q}{1-q}} \\ && \nonumber -\frac{\beta v_y^2\Gamma_y}{\omega_f}\left[1+(q-1)\beta\omega\right]^{\frac{q}{1-q}}-\beta^2v_y^2q\Gamma_y\left[1+(q-1)\beta\omega\right]^{\frac{2q-1}{1-q}}\\ && -\frac{\beta v_yv_z\Gamma_z}{\omega_f}\left[1+(q-1)\beta\omega\right]^{\frac{q}{1-q}}-\beta^2v_yv_zq\Gamma_z\left[1+(q-1)\beta\omega\right]^{\frac{2q-1}{1-q}}
.\ee
Now, substituting the expressions of the above partial derivatives in eq. \eqref{R.B.T.E.(3)} and then simplifying by dropping the higher order velocity terms, we get
\be\label{R.B.T.E.(4)}
-q_fE\tau_fv_x+\left(\Gamma_xv_x+\Gamma_yv_y+\Gamma_zv_z\right)
-\omega_c\tau_f\left(v_x\Gamma_y-v_y\Gamma_x\right)+\tau_f^2\omega_cq_fEv_y=0
,\ee
where $\omega_c=\frac{q_fB}{\omega_f}$ is known as cyclotron frequency. Equating the coefficients of $v_x$, $v_y$ and $v_z$ on both sides of eq. \eqref{R.B.T.E.(4)} 
and then solving, we obtain 
\be
&&\label{Gammax}\Gamma_x=\frac{q_fE\tau_f\left(1-\omega_c^2\tau_f^2\right)}{1+\omega_c^2\tau_f^2}, \\ 
&&\label{Gammay}\Gamma_y=-\frac{2q_fE\omega_c\tau_f^2}{1+\omega_c^2\tau_f^2}, \\
&&\label{Gammaz}\Gamma_z=0
.\ee
Using the above values in the ansatz \eqref{ansatz}, $\delta f_q$ is obtained as
\be\label{deltaf.q}
\delta f_q=\left[2q_fEv_x\beta\left(\frac{\tau_f}{1+\omega_c^2\tau_f^2}\right)-2q_fEv_y\beta\left(\frac{\omega_c\tau_f^2}{1+\omega_c^2\tau_f^2}\right)\right]\left[1+(q-1)\beta(\omega_f-\mu_f)\right]^{\frac{q}{1-q}}
.\ee
Similarly, $\delta \bar{f_q}$ is found out to be
\be\label{deltaf.aq}
\delta \bar{f_q}=\left[2{\bar q_f}Ev_x\beta\left(\frac{\tau_{\bar{f}}}{1+\omega_c^2\tau_{\bar{f}}^2}\right)-2{\bar q_f}Ev_y\beta\left(\frac{\omega_c\tau_{\bar{f}}^2}{1+\omega_c^2\tau_{\bar{f}}^2}\right)\right]
\left[1+(q-1)\beta(\omega_f+\mu_f)\right]^{\frac{q}{1-q}}
.\ee
Substituting the expressions of $\delta f_q$ and $\delta \bar{f_q}$ in eq. \eqref{current} and then comparing with eq. \eqref{Multicomponent structure}, the electrical and Hall conductivities are calculated as
\be\label{E.C.}
\nonumber\sigma_{\rm el} &=& \frac{\beta}{3\pi^2}\sum_f g_f q_f^2\int d{\rm p}~\frac{{\rm p}^4}{\omega_f^2} ~ \left[\frac{\tau_f}{1+\omega_c^2\tau_f^2}\left[1+(q-1)\beta(\omega_f-\mu_f)\right]^{\frac{q}{1-q}}\right. \\ && \left.+\frac{\tau_{\bar{f}}}{1+\omega_c^2\tau_{\bar{f}}^2}\left[1+(q-1)\beta(\omega_f+\mu_f)\right]^{\frac{q}{1-q}}\right], \\ 
\label{H.Conductivity}\nonumber\sigma_{\rm H} &=& \frac{\beta}{3\pi^2}\sum_f g_f q_f^2\int d{\rm p}~\frac{{\rm p}^4}{\omega_f^2} ~ \left[\frac{\omega_c\tau_f^2}{1+\omega_c^2\tau_f^2}\left[1+(q-1)\beta(\omega_f-\mu_f)\right]^{\frac{q}{1-q}}\right. \\ && \left.+\frac{\omega_c\tau_{\bar{f}}^2}{1+\omega_c^2\tau_{\bar{f}}^2}\left[1+(q-1)\beta(\omega_f+\mu_f)\right]^{\frac{q}{1-q}}\right]
.\ee

\subsubsection{Response of the nonextensivity to the heat flow in a weakly magnetized QCD medium}
At finite magnetic field, the Navier-Stokes equation for heat flow takes the following form, 
\begin{eqnarray}\label{Multicomponent structure (2)}
Q^i=-\left(\kappa_0\delta^{ij}+\kappa_1\epsilon^{ijk}b_k+\kappa_2b^ib^j\right)
\left[\partial_j T - \frac{T}{\varepsilon+P}\partial_j P\right] 
,\end{eqnarray}
where $\kappa_0$, $\kappa_1$ and $\kappa_2$ represent different components of heat transport and $\mathbf{b}=\frac{\mathbf{B}}{B}$. If gradients of temperature and pressure are 
orthogonal to the magnetic field, then the third term in the right hand side vanishes 
and thus, eq. \eqref{Multicomponent structure (2)} becomes
\begin{eqnarray}\label{heat2}
Q^i=-\left(\kappa\delta^{ij}+\kappa_{\rm H}\epsilon^{ij}\right)\left[\partial_j T - \frac{T}{\varepsilon+P}\partial_j P\right] 
.\end{eqnarray}
Here $\kappa_0=\kappa$ represents the thermal conductivity and $\kappa_1=\kappa_{\rm H}$ denotes the Hall-type thermal conductivity. To determine these conductivities within the nonextensive Tsallis mechanism, we 
rewrite eq. \eqref{R.B.T.E.} using the ansatz \eqref{ansatz} as
\be\label{eq.1}
L+\beta\left[1+(q-1)\beta\omega\right]^{\frac{q}{1-q}}\left(\Gamma_xv_x+\Gamma_yv_y+\Gamma_zv_z\right)-q_fB\tau_f\left(v_x\frac{\partial f_q^\prime}{\partial p_y}-v_y\frac{\partial f_q^\prime}{\partial p_x}\right)=0
~,\ee
where $L=\frac{\tau_f}{p_0}p^\mu\frac{\partial f_q}{\partial x^\mu}$. We note that, 
for the calculation of heat transport coefficients, we have dropped the electric 
field part from the relativistic Boltzmann transport equation. Since 
magnetic field is taken along z-direction and the gradients of temperature and 
pressure are orthogonal to the magnetic field, no explicit dependence of 
magnetic field on the temperature and pressure gradients along z-direction 
can be observed. Now, $L$ is calculated as
\be\label{L}
\nonumber L &=& \tau_f\beta^2\left[1+(q-1)\beta\omega\right]^{\frac{q}{1-q}}\left(\omega_f-h_f\right)v_x\left(\partial^xT-\frac{T}{nh_f}\partial^xP\right) \\ && \nonumber+\tau_f\beta^2\left[1+(q-1)\beta\omega\right]^{\frac{q}{1-q}}\left(\omega_f-h_f\right)v_y\left(\partial^yT-\frac{T}{nh_f}\partial^yP\right) \\ && +\tau_f\beta\left[1+(q-1)\beta\omega\right]^{\frac{q}{1-q}}\left[p_0\frac{DT}{T}-\frac{p^\mu p^\alpha}{p_0}\nabla_\mu u_\alpha+TD\left(\frac{\mu_f}{T}\right)\right]
,\ee
where $D=u^\mu\partial_\mu$ and $\nabla_\mu=\partial_\mu-u_\mu u_\nu\partial^\nu$. Substituting the values of $L$, $\frac{\partial f_q^\prime}{\partial p_x}$ and 
$\frac{\partial f_q^\prime}{\partial p_y}$ in eq. \eqref{eq.1}, and then 
simplifying by dropping the higher order velocity terms, we have
\be\label{eq.2}
&& \nonumber\beta\left(\omega_f-h_f\right)v_x\left(\partial^xT-\frac{T}{nh_f}\partial^xP\right)+\frac{\Gamma_xv_x}{\tau_f}-\omega_c\Gamma_yv_x \\ && \nonumber+\beta\left(\omega_f-h_f\right)v_y\left(\partial^yT-\frac{T}{nh_f}\partial^yP\right)+\frac{\Gamma_yv_y}{\tau_f}+\omega_c\Gamma_xv_y \\ && +\frac{\Gamma_zv_z}{\tau_f}+p_0\frac{DT}{T}-\frac{p^\mu p^\alpha}{p_0}\nabla_\mu u_\alpha+TD\left(\frac{\mu_f}{T}\right)=0
.\ee
Equating the coefficients of $v_x$, $v_y$ and $v_z$ on both sides of eq. \eqref{eq.2} 
and then solving, we get 
\be
\label{Gamma(x)}\Gamma_x &=& -\frac{\beta\tau_f\left(\omega_f-h_f\right)}{\left(1+\omega_c^2\tau_f^2\right)}\left(\partial^xT-\frac{T}{nh_f}\partial^xP\right) -\frac{\beta\omega_c\tau_f^2\left(\omega_f-h_f\right)}{\left(1+\omega_c^2\tau_f^2\right)}\left(\partial^yT-\frac{T}{nh_f}\partial^yP\right), \\ 
\label{Gamma(y)}\Gamma_y &=& -\frac{\beta\tau_f\left(\omega_f-h_f\right)}{\left(1+\omega_c^2\tau_f^2\right)}\left(\partial^yT-\frac{T}{nh_f}\partial^yP\right) +\frac{\beta\omega_c\tau_f^2\left(\omega_f-h_f\right)}{\left(1+\omega_c^2\tau_f^2\right)}\left(\partial^xT-\frac{T}{nh_f}\partial^xP\right), \\ 
\label{Gamma(z)}\Gamma_z &=& 0
.\ee
Substituting the above values in ansatz \eqref{ansatz}, $\delta f_q$ is calculated as
\be\label{deltaf.q1}
\nonumber\delta f_q &=& -\beta^2\left[1+(q-1)\beta(\omega_f-\mu_f)\right]^{\frac{q}{1-q}}\frac{\tau_f(\omega_f-h_f)}{\left(1+\omega_c^2\tau_f^2\right)}\left[v_x\left(\partial^xT-\frac{T}{nh_f}\partial^xP\right)\right. \\ && \left.\nonumber+v_y\left(\partial^yT-\frac{T}{nh_f}\partial^yP\right)\right]-\beta^2\left[1+(q-1)\beta(\omega_f-\mu_f)\right]^{\frac{q}{1-q}} \\ && \times\frac{\omega_c\tau_f^2(\omega_f-h_f)}{\left(1+\omega_c^2\tau_f^2\right)}\left[v_x\left(\partial^yT-\frac{T}{nh_f}\partial^yP\right)-v_y\left(\partial^xT-\frac{T}{nh_f}\partial^xP\right)\right]
.\ee
Similarly, $\delta \bar{f_q}$ is obtained as
\be\label{deltaf.aq1}
\nonumber\delta \bar{f_q} &=& -\beta^2 \left[1+(q-1)\beta(\omega_f+\mu_f)\right]^{\frac{q}{1-q}}\frac{\tau_{\bar{f}}(\omega_f-\bar{h}_f)}{\left(1+\omega_c^2\tau_{\bar{f}}^2\right)}\left[v_x\left(\partial^xT-\frac{T}{n\bar{h}_f}\partial^xP\right)\right. \\ && \left.\nonumber+v_y\left(\partial^yT-\frac{T}{n\bar{h}_f}\partial^yP\right)\right]-\beta^2\left[1+(q-1)
\beta(\omega_f+\mu_f)\right]^{\frac{q}{1-q}} \\ && \times\frac{\omega_c\tau_{\bar{f}}^2(\omega_f-\bar{h}_f)}{\left(1+\omega_c^2\tau_{\bar{f}}^2\right)}\left[v_x\left(\partial^yT-\frac{T}{n\bar{h}_f}\partial^yP\right)-v_y\left(\partial^xT-\frac{T}{n\bar{h}_f}\partial^xP\right)\right]
.\ee
Using the values of $\delta f_q$ and $\delta \bar{f_q}$ in eq. \eqref{heat1 (1)} 
and then comparing with eq. \eqref{heat2}, the thermal and Hall-type thermal 
conductivities are determined as
\be\label{H.C.}
\nonumber\kappa &=& \frac{\beta^2}{6\pi^2}\sum_f g_f\int d{\rm p}~\frac{{\rm p}^4}{\omega_f^2} ~ \left[\frac{\tau_f}{1+\omega_c^2\tau_f^2}\left(\omega_f-h_f\right)^2\left[1+(q-1)\beta(\omega_f-\mu_f)\right]^{\frac{q}{1-q}}\right. \\ && \left.+\frac{\tau_{\bar{f}}}{1+\omega_c^2\tau_{\bar{f}}^2}\left(\omega_f-\bar{h}_f\right)^2
\left[1+(q-1)\beta(\omega_f+\mu_f)\right]^{\frac{q}{1-q}}\right], \\ 
\label{H.C.(1)}\nonumber\kappa_{\rm H} &=& \frac{\beta^2}{6\pi^2}\sum_f g_f\int d{\rm p}~\frac{{\rm p}^4}{\omega_f^2} ~ \left[\frac{\omega_c\tau_f^2}{1+\omega_c^2\tau_f^2}\left(\omega_f-h_f\right)^2\left[1+(q-1)\beta(\omega_f-\mu_f)\right]^{\frac{q}{1-q}}\right. \\ && \left.+\frac{\omega_c\tau_{\bar{f}}^2}{1+\omega_c^2\tau_{\bar{f}}^2}\left(\omega_f-\bar{h}_f\right)^2\left[1+(q-1)
\beta(\omega_f+\mu_f)\right]^{\frac{q}{1-q}}\right]
.\ee

The above analysis on the charge and heat conductivities is studied by 
considering the quasiparticle or thermal masses of particles within 
the quasiparticle model. Partons acquire thermal masses due to their 
interactions with the surrounding thermal medium. In a hot and dense 
QCD medium, the quasiparticle mass (squared) of quark up to one-loop 
is given \cite{Braaten:PRD45'1992,Peshier:PRD66'2002} by
\be\label{Q.P.M.}
m_{fT}^2=\frac{g^2T^2}{6}\left(1+\frac{\mu_f^2}{\pi^2T^2}\right)
.\ee
We note that, all flavors are assigned with the same chemical potential, 
{\em i.e.} $\mu_f=\mu$. 

\section{Results and discussions}
\begin{figure}[]
\begin{center}
\includegraphics[width=13.9cm]{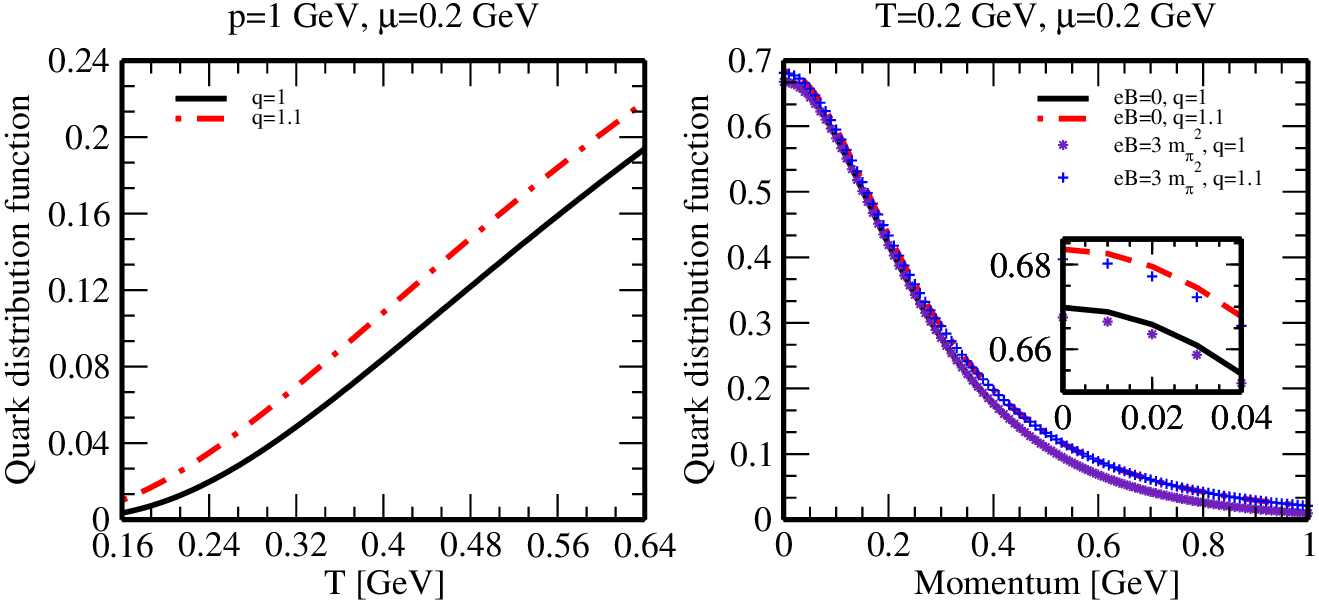}
\caption{Variation of quark distribution function with temperature (left panel) and with momentum (right panel).}\label{dftu}
\end{center}
\end{figure}

In kinetic theory, the particle distribution functions embody most of the information on the transport coefficients and thus a study of the effect of nonextensivity on distribution function is first 
carried out before discussing its effect on the transport coefficients. Figure \ref{dftu} shows the 
$u$ quark distribution function in terms of temperature and momentum at a fixed chemical potential for different values of the nonextensive parameter and magnetic field. In particular, the left panel of figure \ref{dftu} depicts the quark distribution function in terms of temperature at a fixed momentum and the right panel of figure \ref{dftu} shows the same in terms of momentum at a fixed temperature. It is observed that the values are higher for the Tsallis distribution ($q=1.1$) when compared with the Fermi-Dirac distribution ($q=1$) over the shown range of temperature and the difference is not significant at low temperatures (left panel). On the other hand, the difference between the Tsallis distribution and the Fermi-Dirac distribution is exiguous at low momenta (right panel). Although the two types of distribution functions get slightly reduced in the presence of weak magnetic field as compared to zero magnetic field, the strength of Tsallis distribution function is always higher than that of the Fermi-Dirac distribution function. 

We note that, in this work, the temperature and chemical potential are treated independent of $q$, 
{\em i.e.} $T\neq T(q)$ and $\mu\neq \mu(q)$. Thus, we plot the transport coefficients as functions 
of temperature at different $q$ values, which gives the information on how the transport coefficients get affected by the nonextensivity when the temperature of the thermal medium changes. On the other hand, the thermal distribution functions of particles depend on all the aforesaid parameters, {\em i.e.} $T$, $\mu$ and $q$. How the distribution function depends on $q$ is described above. Similarly, the energy-momentum tensor ($T^{\alpha\beta}$) and the particle flow four-vector ($N^\alpha$) depend on $q$ through the particle distribution function, thus the energy density ($\varepsilon=u_\alpha T^{\alpha\beta} u_\beta$) as well as the particle number density ($n=N^\alpha u_\alpha$) also depend on $q$ parameter, in addition to their dependence on $T$ and $\mu$. 

If one keeps the energy density or the particle number density at the same value in nonextensive Tsallis distribution ($q>1$) and in Boltzmann type distribution ($q=1$), then the Tsallis distribution, 
as compared to the Boltzmann one, leads to smaller values of $T$, {\em i.e.} with increasing $q$ value, $T$ decreases. For example, the references \cite{Cleymans:JPG36'2009,Wilk:EPJA40'2009} had shown that, in order to keep the hadron yields (in high energy heavy ion collisions) the same in both nonextensive Tsallis and Boltzmann distributions, $T$ is adjusted to lower values for increasing $q$ value. Since the aim of the present paper is to see how the nonextensivity ($q>1$) affects the transport coefficients of the hot and dense QCD matter in kinetic theory approach for a range of temperature and chemical potential, we have not considered the correlation between $q$, $T$ and $\mu$. 

\subsection{Response functions of charge flow: electrical and Hall conductivities}
\begin{figure}[]
\begin{center}
\begin{tabular}{c c}
\includegraphics[width=6.3cm]{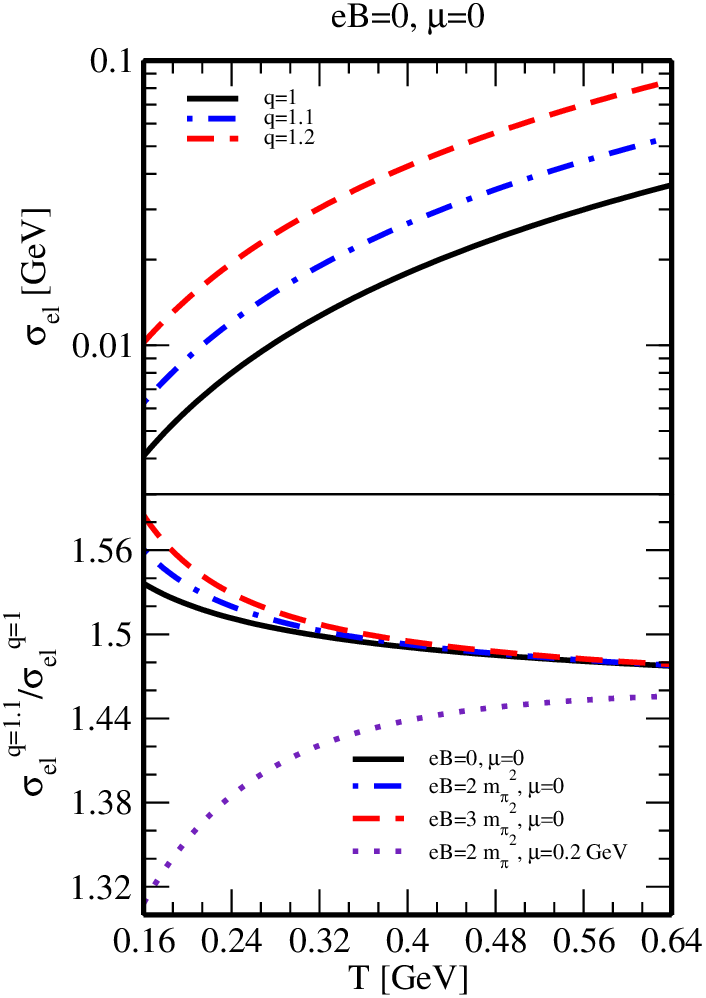}&
\hspace{0.74 cm}
\includegraphics[width=6.3cm]{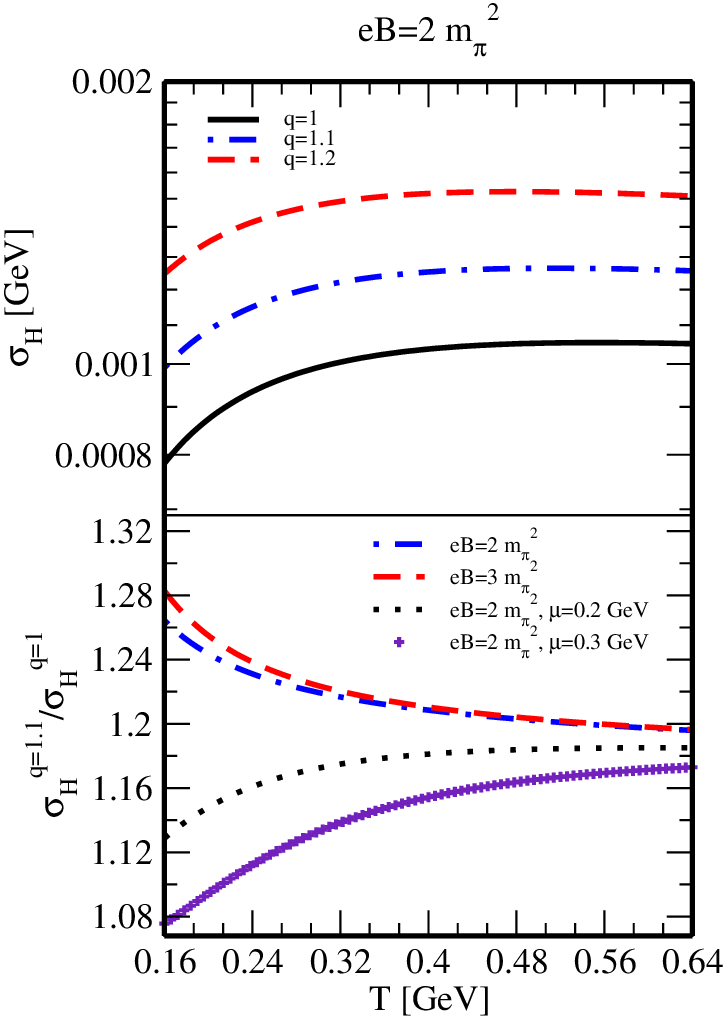} \\
a & b
\end{tabular}
\caption{Variations of (a) electrical conductivity and (b) Hall conductivity with temperature for different values of the nonextensive parameter at weak magnetic field and 
finite chemical potential.}\label{Fig.1}
\end{center}
\end{figure}

Figure \ref{Fig.1} depicts the variations of electrical ($\ec$) and Hall ($\eh$) conductivities of the hot QCD matter with temperature for different values of $q$. In particular, the upper panels of figures \ref{Fig.1}a and \ref{Fig.1}b show an enhancement in their magnitudes when $q$ is slightly above unity and this increase is almost uniform over the entire range of temperature, thus it indicates that the nonextensivity facilitates the charge transport in hot QCD matter. The lower panels of figures \ref{Fig.1}a and \ref{Fig.1}b show the variations of the ratios of $\ec$ and $\eh$ at $q=1.1$ and at $q=1$ with $T$ for different conditions of magnetic field and chemical potential. The nonextensive $\ec$ as well as $\eh$ get deviated from the corresponding thermally equilibrated values with the increase of magnetic field, whereas, the emergence of finite chemical potential brings $\ec$ and $\eh$ a bit closer to their equilibrated values. As compared to the Hall conductivity, the effect of the nonextensivity is more evident on the electrical conductivity in the weak magnetic field regime as $\ec$ is the dominant contribution and $\eh$ is a weak contribution of the charge transport in the said regime. 

\subsection{Response functions of heat flow: thermal and Hall-type thermal conductivities}
\begin{figure}[]
\begin{center}
\begin{tabular}{c c}
\includegraphics[width=6.3cm]{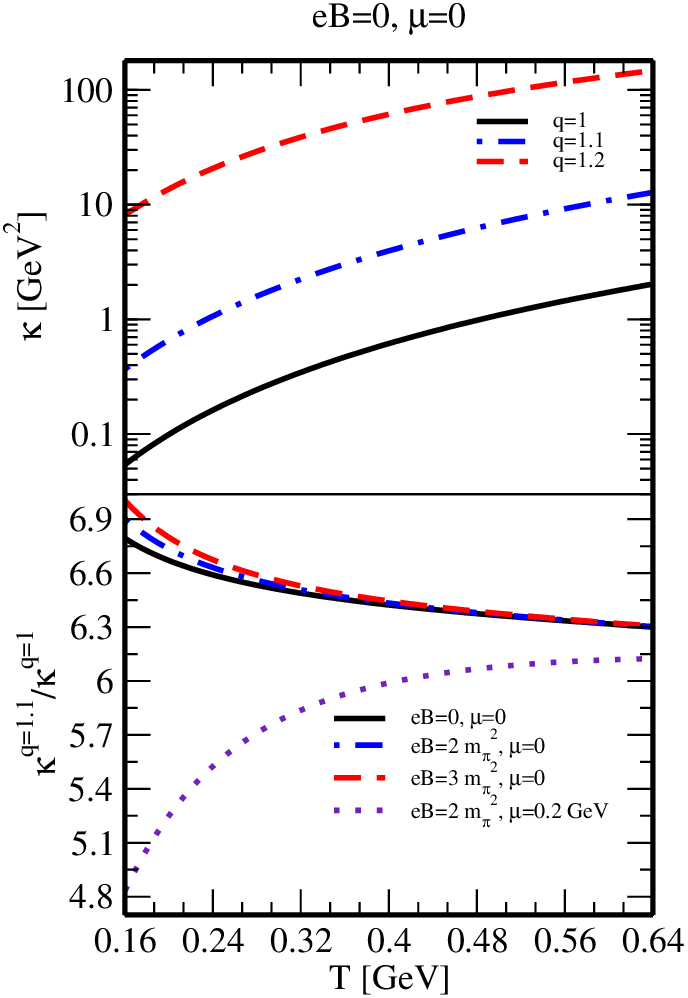}&
\hspace{0.74 cm}
\includegraphics[width=6.3cm]{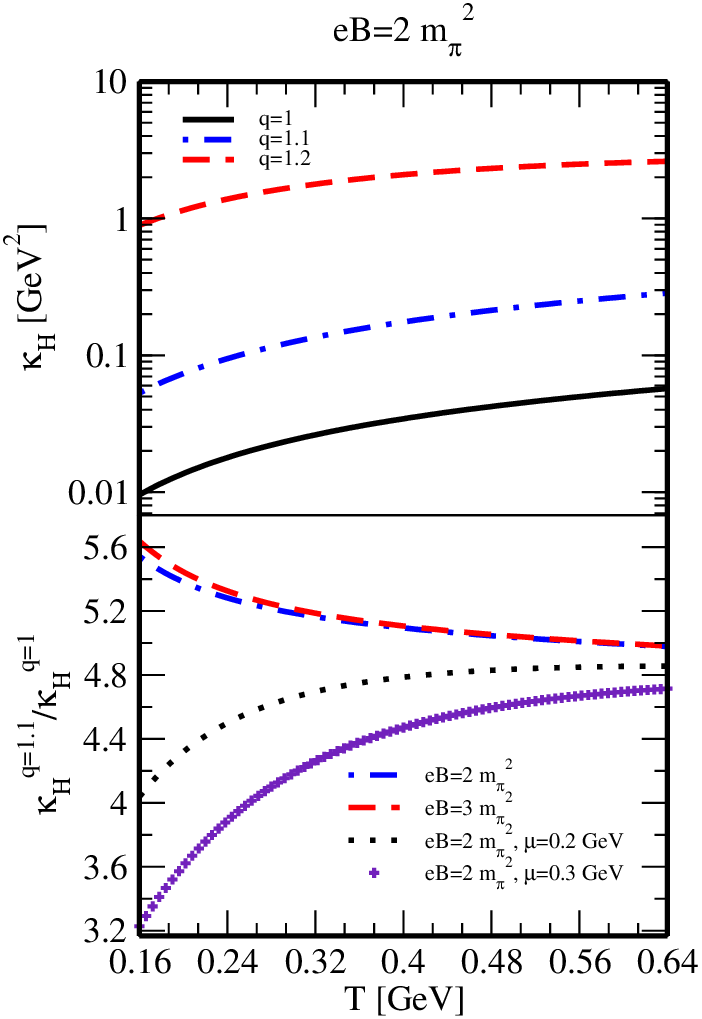} \\
a & b
\end{tabular}
\caption{Variations of (a) thermal conductivity and (b) Hall-type thermal conductivity with temperature for different values of the nonextensive parameter at 
weak magnetic field and finite chemical potential.}\label{Fig.2}
\end{center}
\end{figure}

The thermal ($\kappa$) and Hall-type thermal ($\ek$) conductivities of the hot QCD matter are shown as functions of temperature for different values of $q$ in figure \ref{Fig.2}. From the upper panels of figures \ref{Fig.2}a and \ref{Fig.2}b, one can see that both $\kappa$ and $\ek$ increase with an increase of $q$. Thus, the nonextensivity amplifies the heat transport in hot QCD matter. Further, from the lower panels of figures \ref{Fig.2}a and \ref{Fig.2}b, it is observed that the presence of magnetic field increases the deviation of non-equilibrium $\kappa$ and $\ek$ from their respective thermally equilibrated values, unlike the case of finite chemical potential which decreases the deviation. Just like the charge transport case, the effect of nonextensivity is less pronounced for $\ek$, because in the weak magnetic field regime, thermal conductivity is the dominant contribution of the heat transport. At high temperatures, the extra deviation occurred due to the magnetic field gets waned. 

\section{Observables}
In this section, a detailed study on the Knudsen number and its association with elliptic
flow coefficient is carried out. These quantities decipher the local equilibrium
property of the medium and the extent of interactions among the produced particles in
heavy ion collisions. 

The Knudsen number is defined as the ratio of the mean free path ($\lambda$) to the 
characteristic length scale ($l$) of the medium, {\em i.e.} $\Omega={\lambda}/{l}$. 
For an equilibrium system, $l$ should be larger than $\lambda$. Since $\lambda={3\kappa}/{(vC_V)}$, the Knudsen number can be rewritten as
\be\label{Knudsen number}
\Omega=\frac{3\kappa}{lvC_V}
~,\ee
where $v$ and $C_V$ denote the relative speed and the specific heat at constant 
volume, respectively. The Knudsen number quantifies the 
degree of separation between the microscopic and macroscopic 
length scales of the system. For the applicability of equilibrium 
hydrodynamics, these two length scales require to be 
sufficiently separated, {\em i.e.} $\Omega$ must be very small or 
less than unity. In this analysis, we have used $v\simeq 1$ and $l=4$ fm, 
and determined $C_V$ from the energy-momentum tensor through the relation, 
$C_V=\partial (u_\mu T^{\mu\nu}u_\nu)/\partial T$. 

The elliptic flow ($v_2$) reflects the anisotropy which originates from the overlapping region of nuclei 
and measures the extent of interactions/reinteractions between the particles produced in the noncentral 
heavy ion collisions. The initial asymmetry in the geometry of the matter distribution manifests the 
azimuthal anisotropy, which gets converted into the momentum anisotropy, thus contributing towards the emergence of elliptic flow \cite{Ollitrault:PRD46'1992}. The deviation from equilibrium can decrease the magnitude of this flow and if the collisions are frequent enough, then the system drives towards local equilibrium. Thus, the elliptic flow provides the information about the onset of thermalization in heavy ion collisions. The elliptic flow can be defined in terms of the Knudsen number \cite{Bhalerao:PLB627'2005,Drescher:PRC76'2007,Gombeaud:PRC77'2008} as
\be\label{Elliptic flow}
v_2=\frac{v_2^h}{1+\frac{\Omega}{\Omega_0}}
,\ee
where $v_2^h$ denotes the elliptic flow in the hydrodynamic limit 
($\Omega\rightarrow 0$ limit) and $\Omega_0$ is the value 
of the Knudsen number obtained by observing the transition 
between the hydrodynamic regime and the free streaming particle 
regime. According to the transport calculation in 
ref. \cite{Gombeaud:PRC77'2008}, $\Omega_0 \approx 0.7$ and $v_2^h \approx 0.1$. Experimentally observed values of $v_2$ at RHIC \cite{Masui:NPA774'2006,Wang:NPA774'2006} are found to be larger than the theoretical estimates due to the strongly interacting nature of quark gluon plasma. As per the Cu-Cu PHOBOS collaboration \cite{Manly:NPA774'2006}, the reason for this large magnitude is mainly attributed to the fluctuations in the nucleon positions, which get transmitted into the fluctuations in the almond shape, thus stemming larger values of $v_2$ \cite{Bhalerao:PLB641'2006}. The presence of external fields could also influence the elliptic flow, {\em e.g.} an enhancement in the elliptic flow due to the presence of magnetic fields had been observed in references \cite{Mohapatra:MPLA26'2011,Rath:EPJA59'2023}. 

\begin{figure}[]
\begin{center}
\begin{tabular}{c c}
\includegraphics[width=6.3cm]{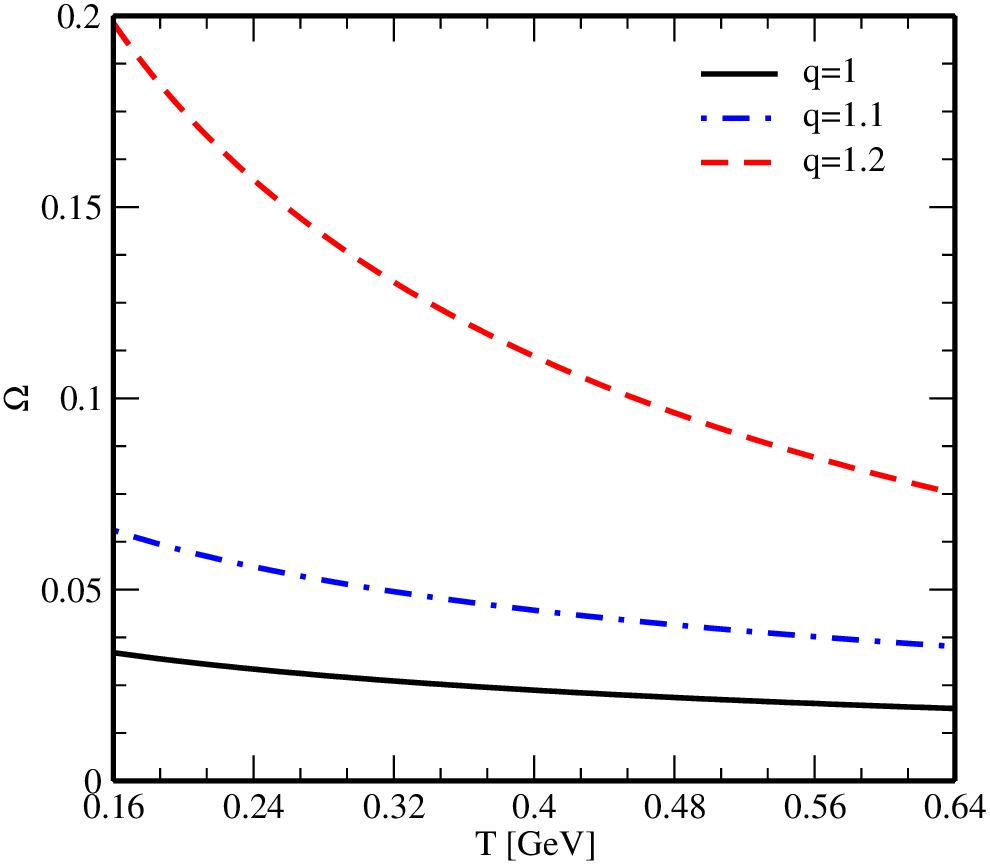}&
\hspace{0.74 cm}
\includegraphics[width=6.3cm]{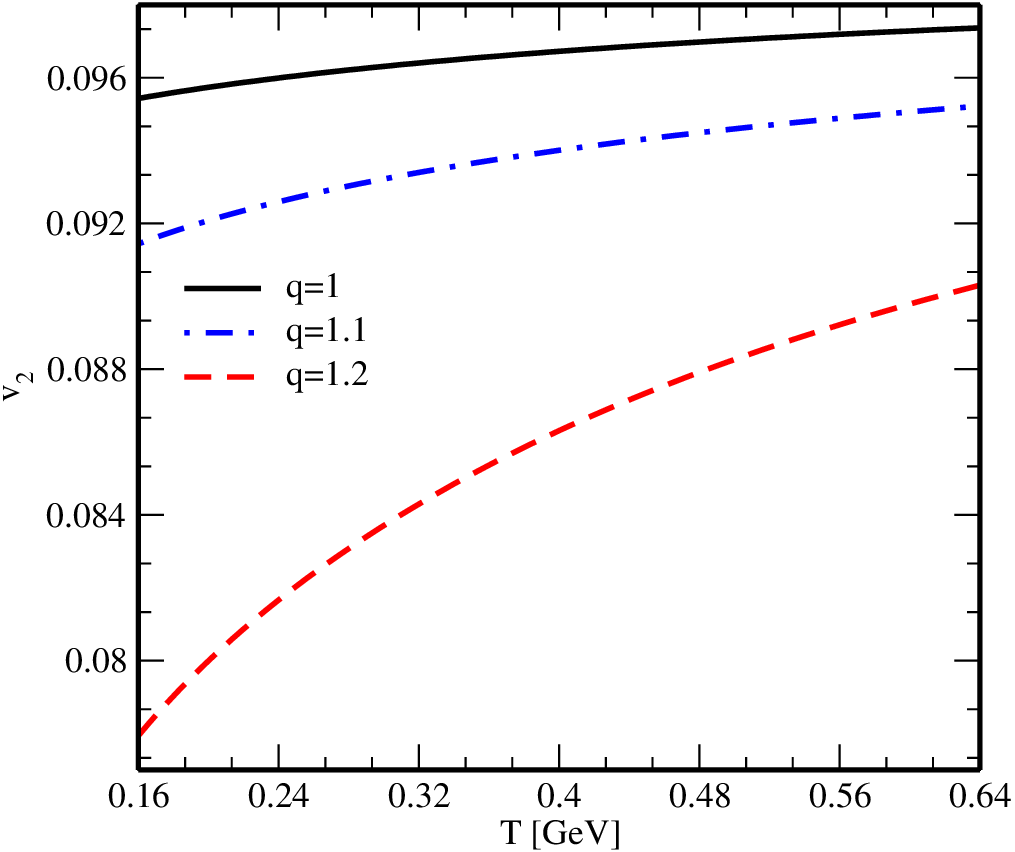} \\
a & b
\end{tabular}
\caption{Variations of (a) the Knudsen number and (b) the elliptic flow with temperature for different values of the nonextensive parameter.}\label{Fig.3}
\end{center}
\end{figure}

\begin{figure}[]
\begin{center}
\begin{tabular}{c c}
\includegraphics[width=6.3cm]{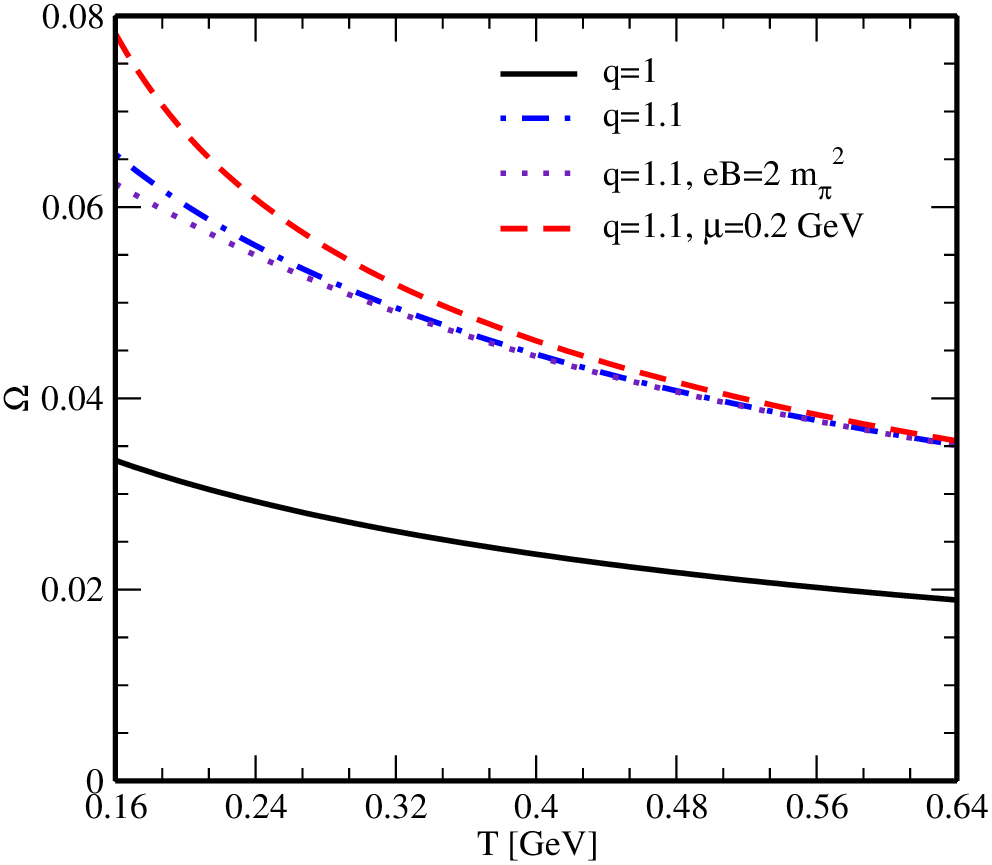}&
\hspace{0.74 cm}
\includegraphics[width=6.3cm]{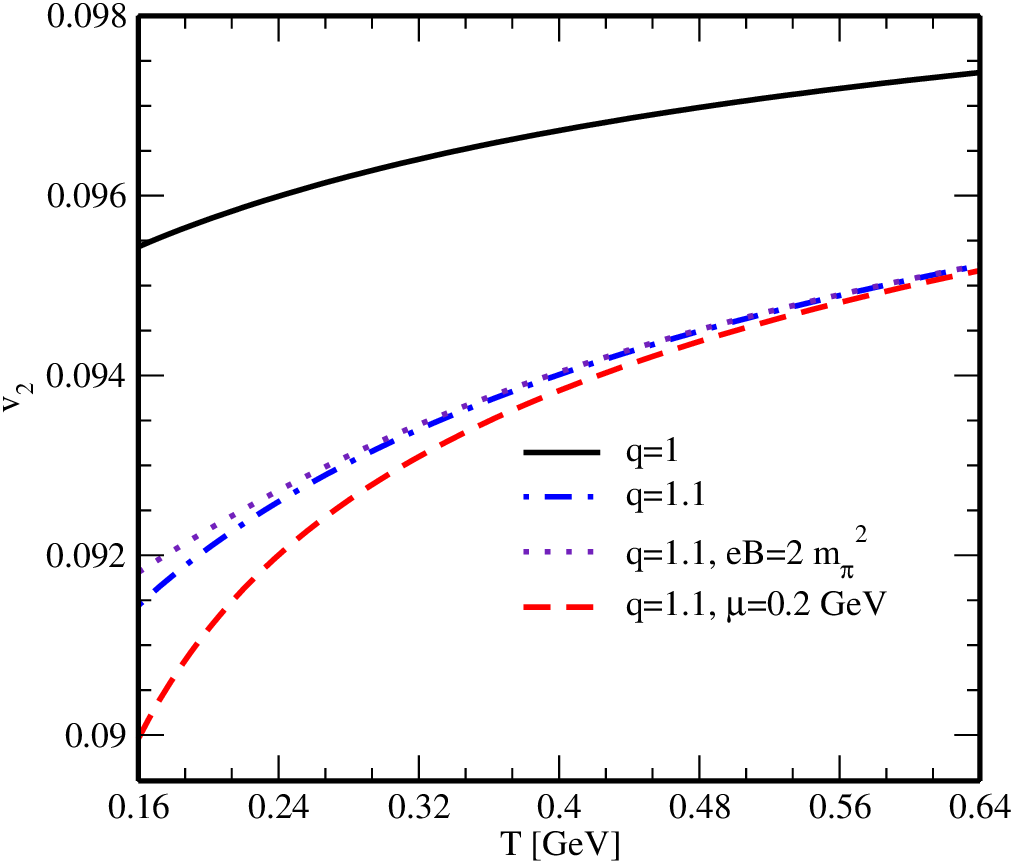} \\
a & b
\end{tabular}
\caption{Variations of (a) the Knudsen number and (b) the elliptic flow with temperature for different values of the nonextensive parameter at finite magnetic field and chemical potential.}\label{Fig.4}
\end{center}
\end{figure}

\begin{figure}[]
\begin{center}
\includegraphics[width=6.3cm]{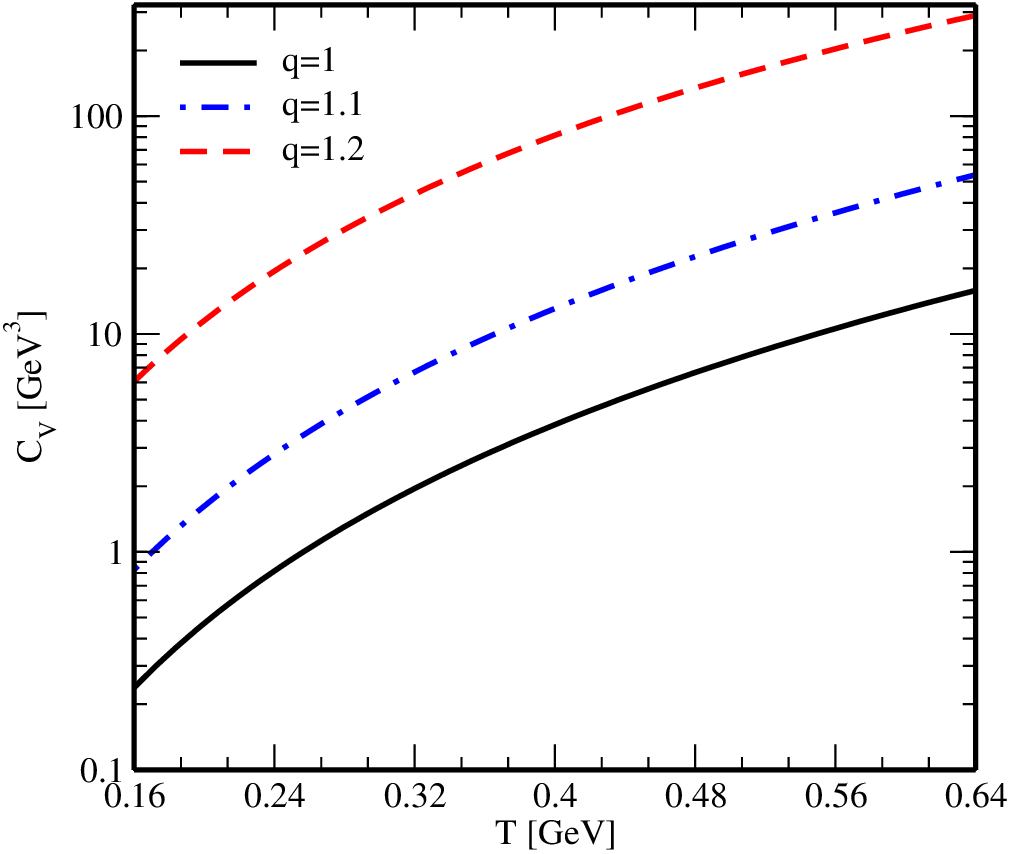}
\caption{Variation of the specific heat with temperature for different values of the nonextensive parameter.}\label{Fig.5}
\end{center}
\end{figure}

Figures \ref{Fig.3}a and \ref{Fig.3}b show how the nonextensivity affects the Knudsen number and the elliptic flow, respectively. With the increase of $q$, the Knudsen number is found to increase, whereas the elliptic flow gets decreased as compared to their counterparts at $q=1$. The increasing behavior of $\Omega$ with $q$ is corroborated by the increase of both $\kappa$ (upper panel of figure \ref{Fig.2}a) and $C_V$ (figure \ref{Fig.5}), with the increase of the former one being larger than that of the latter one. The opposite behaviors of $\Omega$ and $v_2$ can be elucidated from eq. \eqref{Elliptic flow}, which shows that both $\Omega$ and $v_2$ are almost inversely proportional to each other. The increase of $\Omega$ with $q$ describes that the mean free path approaches towards the characteristic length scale of the medium, thus taking the system a bit away from the local equilibrium state. This agrees with the fact that the deviation of $q$ from unity drives the system towards its nonequilibrium state. As $q$ increases, there is a decrease of the number of collisions, which results into a smaller anisotropic flow, hence $v_2$ gets waned. The reduction of elliptic flow with the enhancement of the nonextensivity can also be realized from the enhancement of thermal conductivity (through its dependence on mean free path) in the similar environment (figure \ref{Fig.2}a). 

Figures \ref{Fig.4}a and \ref{Fig.4}b respectively depict the effects of magnetic field and chemical 
potential on the Knudsen number and the elliptic flow for a thermal medium with 
finite nonextensivity. From the above discussion (see figures \ref{Fig.2}a, \ref{Fig.3}a, \ref{Fig.3}b and \ref{Fig.5}), it is evident that the Knudsen number gets enhanced and the elliptic flow becomes reduced as $q$ 
changes from 1 to 1.1. Now from figure \ref{Fig.4}a, it can be seen that the presence of magnetic 
field pushes the Knudsen number towards its value at $q=1$ (black solid line), contrary to the 
chemical potential which takes it further away from the said value. From figure \ref{Fig.4}b, it is 
inferred that the emergence of magnetic field shifts the elliptic flow towards its value at $q=1$ 
(black solid line), whereas chemical potential further deviates it away from this value. The main 
reason behind the opposite effects of magnetic field and chemical potential on aforesaid observables 
is attributed to their opposite effects on the thermal conductivity (lower panel of figure \ref{Fig.2}a). 

\section{Conclusions}
In this work, we focused on the effects of the nonextensivity on the conduction of charge and heat in hot and dense QCD matter at finite magnetic field. The electrical conductivity ($\ec$) and the thermal conductivity ($\kappa$) were calculated using the relativistic Boltzmann transport equation in the kinetic theory approach within the nonextensive Tsallis formalism. The effect of the finite magnetic field was also studied. The presence of magnetic field additionally introduced two transport coefficients, namely, the Hall conductivity ($\eh$) and the Hall-type thermal conductivity ($\ek$), which were also studied using the nonextensive framework. The lifetime of magnetic field was observed to increase with an increase of the nonextensive parameter. The transport coefficients ($\ec$, $\eh$, $\kappa$ and $\ek$) were observed to increase with the increase of the nonextensivity of the medium. This could have an observable effect on some observables associated with the aforesaid transport coefficients, such as the Knudsen number, the elliptic flow etc. Since these observables carry the information about the local equilibrium property of the matter, interactions between produced particles in heavy ion collisions etc., one can comprehend how the nonextensivity influences them. 

\section{Acknowledgments}
One of us (S. R.) would like to thank the Indian Institute of Technology Bombay 
for the Institute postdoctoral fellowship and S. D. acknowledges the SERB Power 
Fellowship, SPF/2022/000014 for the support on this work.

\end{document}